# Cellular morphogenesis of three-dimensional tensegrity structures

Omar ALOUI[a], Jessica FLORES[b], David ORDEN[c], Landolf RHODE-BARBARIGOS*

[a] Department of Civil, Architectural & Environmental Engineering, University of Miami, 1251 Memorial Drive, Coral Gables, FL 33146-0630, USA, Email: omar.aloui@miami.edu

[b] Department of Civil, Architectural & Environmental Engineering, University of Miami, 1251 Memorial Drive, Coral Gables, FL 33146-0630, USA, Email: j.flores7@umiami.edu

[c] Departamento de Física y Matemáticas, Universidad de Alcalá. Ctra. Madrid-Barcelona, Km. 33,600, 28805, Alcalá de Henares, Spain, Email: david.orden@uah.es

* Department of Civil, Architectural & Environmental Engineering, University of Miami, 1251 Memorial Drive, Coral Gables, FL 33146-0630, USA, Email: landolfrb@miami.edu

**Abstract**

The topology and form finding of tensegrity structures have been studied extensively since the introduction of the tensegrity concept. However, most of these studies address topology and form separately, where the former represented a research focus of rigidity theory and graph theory, while the latter attracted the attention of structural engineers. In this paper, a biomimetic approach for the combined topology and form finding of spatial tensegrity systems is introduced. Tensegrity cells, elementary infinitesimally rigid self-stressed structures that have been proven to compose any tensegrity, are used to generate more complex tensegrity structures through the morphogenesis mechanisms of adhesion and fusion. A methodology for constructing a basis to describe the self-stress space is also provided. Through the definition of self-stress, the cellular morphogenesis method can integrate design considerations, such as a desired shape or number of nodes and members, providing great flexibility and control over the tensegrity structure generated.

## 1. Introduction

### 1.1. Definitions and applications

Tensegrity structures are reticulated prestressed free-standing structures in a state of self-equilibrium composed of members in tension and members in compression. The term was first used by Buckminster Fuller in 1962 to describe Kenneth Snelson's sculptures [1,2]. Nowadays, tensegrity definitions vary according to the field. In architecture and engineering [2], tensegrity describes "*a system in stable self-equilibrated state comprising a discontinuous set of compressed components inside a continuum of tensioned components*", while in mathematics and rigidity theory, tensegrity is defined as a self-stressed framework. A framework $T(G,P)$ is a realization of a graph $G(V,E)$ in $d$-space described by its set of vertices $V$ and edges $E$, and a configuration $P=[p_1;p_2;…;p_n]$ (a collection of points described by $d$ coordinates) [3]. In this study, the mathematical definition is followed, with the term geometry referring to the nodal coordinates of the structure while topology refers to the set of members $E$ that link the nodes (connectivity).

Tensegrity quickly evolved from an art concept to a structural system and scientific model that enables the design of systems with a high strength-to-mass ratio [4] and the ability to change shape and/or deploy while maintaining equilibrium [5]. The concept has thus generated interest from mathematicians [3,6], architects [7,8], material [9], structural [10,11], aerospace [12,13], robotics [14,15] and biomechanical engineers [16,17,18]. However, tensegrity structures cannot be fully exploited unless the effects of geometry and topology are properly understood and integrated in the design process.

### 1.2. Topology and form finding of tensegrity structures

The first step in the design process of tensegrity structures is form finding: the process of finding a stable equilibrium configuration for a structure under specific loading and boundary conditions starting from an arbitrary geometry [19]. In tensegrity, the resulting configuration is obtained under prestress with only rigid body motions constrained. Force density [20,21,22] and dynamic relaxation [23,24] are two well-known methods employed often for the form finding of tensegrity structures. However, both methods require a predefined topology and typology of elements as input, and do not control the self-stress and equilibrium geometry of the resulting structure. Tensegrity studies are thus typically based on systems with known topologies and geometries, as identifying a topology that results in a stable tensegrity structure is a challenging task of combinatorial nature. Consequently, any attempt of solving this problem using brute force methods is computationally expensive. Geometric constraints and heuristics have thus been used. Nishimura and Murakami [25] used symmetry considerations to analyze and find the initial shape of cyclic frustum tensegrity modules. Zhang et al. [26] employed element directions, symmetry properties and some nodal positions as constraints to solve the equilibrium problem and identify stable tensegrity structures. The proposed method is however not practical for large irregular systems, as the number of constraints required increases with the number of unknowns. Lee and Lee [27] combined the force-density method with genetic algorithms so that the form-finding process requires no knowledge about the topology of the structure by removing unnecessary tension elements for a selected set of compression elements. However, the method does not provide any control or insight on the resulting tensegrity structure.

The topology search for stable rigid tensegrities, and frameworks in general, has been studied in rigidity theory and graph theory starting with Maxwell counting rules for the static determinacy of structures, and later Laman who provided a full characterization of minimally generically rigid frameworks in the plane [28]. These results were generalized by Recski [29] to apply specifically to self-stressed tensegrity structures: "*A simple graph G with n vertices and 2n-2 edges is generically rigid tensegrity in the plane if and only if $|E'| \leq 2|V'| - 3$ holds for every proper subgraph G'(V',E') of G with at least two vertices*". However, the characterization of three-dimensional minimally rigid tensegrity structures remains an long standing open problem in rigidity theory. de Guzmán and Orden [6] characterized tensegrity topologies through their decomposition into elementary stable topological units. Aloui et al (2018) [30] proposed a generative method for the design of planar tensegrity structures based on the elementary stable topological units defined by de Guzmán and Orden [6]. However, the method cannot be directly extended to three-dimensional structures, as the constitutive topological units for the planar and the spatial case differ and the number of topological combinations to be considered increases.

Obtaining a valid tensegrity topology has been the focus of many other studies. Connelly and Back [31], Connelly and Terrell [32], Masic et al [33] and Sultan et al. [34] used common group-theoretic symmetry property to find structures with a predefined symmetry. The stability conditions of such tensegrities were studied using group representation theory by Zhang and Ohsaki [35,36]. However, these methods require the symmetry properties to be fixed in advance. Rieffel et al. [37] addressed the problem using grammar-based representation graphs which allowed them to find asymmetric irregular structures. Ehara and Kanno [38] tried to broaden the solution space to include irregular tensegrity topologies using mixed integer linear programming (MILP). The method was refined by Xu et al. [39] to include mixed linear quadratic programming (MIQP), which allowed to find class k (k>1) tensegrity structures (tensegrity systems with a maximum of k interconnected compressive members [40]). Lee and Lee [27] combined force density method and genetic algorithms to find these topologies; however, the method requires the nodal positions to be known in advance. Li et al [41] studied the construction of tensegrity structures from one-bar units. Although these methods allow for the identification of topologies for tensegrity structures, most of them apply restrictions on the solution space and they do not provide control over the self-stress in the structure.

Cellular morphogenesis of tensegrity structures represents a bio-inspired approach for the generative design of tensegrity structures that combines topology and geometry finding of tensegrity structures, allowing one to find stable structures with predefined shapes and a predefined number of self-stress states. The method is inspired by the morphogenesis mechanisms of biological cells, providing an intuitive approach to understand the interactions between topology and geometry in tensegrity structures. The idea of mimicking cellular mechanisms for the analysis and design of structures echoes back to the work of Motro [1] who referred to composing modules of complex tensegrity structures as cells, and the work of Canyurt and Hejela [42] who proposed a cellular framework for structural analysis and optimization. Zhang et al. [43] employed a stiffness-matrix-based form-finding method to rapidly find complex tensegrity structures constructed using repetition of the same module. However, in this paper the cell idea is perceived differently with cells being predefined topological entities that compose complex tensegrity structures with the nature of their composition defining the self-equilibrium in the resulting system. The paper starts thus by describing the foundations of cellular morphogenesis. Section 2 presents the mathematical background necessary to understand the principles of the method, along with a description of the tensegrity units (cells) that are employed to compose complex tensegrity structures. Section 3 describes the cellular morphogenesis mechanisms and their implications on the geometry of the structure and its self-stress space. Section 4 focuses on the implementation of the method, while Section 5 presents a series of examples analyzed through the cellular morphogenesis principles.

## 2. Theoretical foundations

### 2.1. Self-equilibrium in tensegrity structures

Tensegrity structures are systems in a state of self-equilibrium [2]. The self-equilibrium is defined by a set of internal forces that depend on the topology and the geometry of the structure. For some topologies, the self-equilibrium is independent of the nodal configuration. These topologies are referred to in rigidity theory as generically rigid graphs. However, for non-generically rigid graphs, the nodal positions have to satisfy specific geometrical conditions in order for the self-equilibrium to exist. This can be seen through the sufficient conditions for the stability of tensegrity structures proposed by Connelly [44] where the third condition states that the member directions do not lie on the same conic at infinity. Although, as will be seen in the remainder of the paper, tensegrity structures can be designed to have underlying generically rigid graphs, the majority of the popular tensegrity systems, such as the Triplex and the Icosahedron, fall into the second category.

Although rigidity theory provides a combinatorial solution for the stability and the existence of a self-equilibrium, in many studies an answer is sought through the solution of the equilibrium equations. Let $E$ be the set of members of the structure and $|E|$ the total number of members. Let $V$ be the set of nodes of the structure and $|V|$ the total number of nodes. Let $\bar{x}_i$ be the vector of Cartesian coordinates in space of node $i$. Considering $w_{ij}$ as a scalar representing the force density (force in the element over the length of the element) or self-stress component of the element linking node $i$ to $j$, the equilibrium at node $i$ of the structure is given by (Pellegrino 1990) [45]:

$$\sum_{\substack{j \\ (i,j) \in E}} (\bar{x}_i - \bar{x}_j) w_{ij} = 0 \qquad (1)$$

The equilibrium at every node results in a system of $3|V|$ equations that can be described algebraically as:

$$Aw = 0 \qquad (2)$$

where $w$ is a vector of $|E|$ self-stress components and $A$ is the equilibrium matrix. Self-stress can thus be defined as the set of force-density values that induce a state of self-equilibrium in the structure without considering external loads or supports, which is characterized algebraically by the null space of the equilibrium matrix $A$:

$$W = nullspace(A) \qquad (3)$$

where $W$ is a basis of the self-stress space.

### 2.2. Mathematical foundation

The method proposed is based on a series of statements, theorems and propositions, adapted from rigidity theory and graph theory. The first statement is a theorem developed by de Guzmán and Orden (2006) [6] focusing on the decomposition of $d$-dimensional tensegrity structures into elementary units, whose three-dimensional case is as follows:

> **Theorem.** Let $T(P)$ be the tensegrity structure defined by the framework $(V,E,P)$ where $G=(V,E)$ is the abstract graph on the set of vertices $V$ and the set of edges $E$, and $P$ is a configuration of points in a three-dimensional space in general position with no four points lying on the same hyperplane. The tensegrity structure $T(P)$ is then a finite sum of elementary units defined by the complete graph on five points, denoted as $K_5$.

The theorem suggests that these complete graphs which have all pairs of vertices connected by an edge, named tensegrity cells in this paper, can be used as building blocks for any tensegrity structure regardless of its topology and geometry.

The second statement characterizes combinatorically the dimension of the self-stress space. In rigidity theory, the dimension of the self-stress space $|W|$ is related to the number of degrees of freedom of the framework $G(P)$, denoted $df$ by the Proposition below proposed by Graver et al. (1993) [46] and adapted here for the three-dimensional space:

> **Proposition.** Let $G(P)$ be a framework in general position $P$ in dimension 3 with $G=(V,E)$ the underlying abstract graph of the framework and $|W|$ the dimension of its self-stress space. The number of degrees of freedom $df$ of the framework $G(P)$ is given by:
>
> $$df = \begin{cases} |W| - (6 + |E| - 3|V|) & \text{if } |V| \geq 3 \\ \dfrac{|V|(|V|-1)}{2} - |E| & \text{if } |V| \leq 4 \end{cases}$$

The proposition reflects a generalization of the Maxwell counting rule for the static and kinematic determinacy of trusses, with the second part of the difference in the first row being the Laman bound:

$$B = 6 + |E| - 3|V| \qquad (4)$$

## 2.3. Three-dimensional tensegrity cells

In this study, any tensegrity structure that has only one self-stress state is referred to as a unicellular organism. If the structure has one self-stress state and its underlying graph is a complete graph $K_5$ on five nodes, then it is called a cell. The two possible configurations of three-dimensional tensegrity cells are illustrated in Figure 1. Although topologically the two cells are the same, the embedding of the abstract graph $K_5$ in the space results in two different structures according to element typology. Elements in both cell types can be classified into two groups of the same type (bars or cables): a group of six elements and a group of four elements. The difference between the two systems resides in the fact that the elements of the four-element group in Type II cells are incident to the same central node, where in Type I cells the elements form a central triangle $P_1P_2P_3$ with the fourth element linking the remaining nodes. In Figure 1, element groups are distinguished using red and blue lines. However, it should be noted that the type of elements is not assigned at this stage as groups can take compression or tension, resulting in four total different structures (duality).

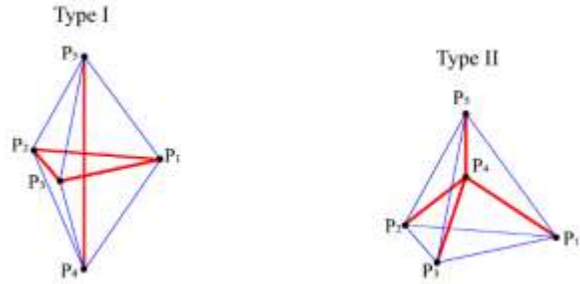

Figure 1: Illustration of three-dimensional tensegrity cells. Element groups are distinguished using red and blue lines.

The cells have one stable self-stress state and are infinitesimally rigid. Cellular morphogenesis exploits the uniqueness of the self-stress state solution in cells to construct a basis for the description of the self-stress in any tensegrity structure composed of cells. The section below describes the development of the analytical solution for the self-stress state in tensegrity cells.

Let $P_1(p_{11},p_{12},p_{13})$, $P_2(p_{21},p_{22},p_{23})$, $P_3(p_{31},p_{32},p_{33})$, $P_4(p_{41},p_{42},p_{43})$ and $P_5(p_{51},p_{52},p_{53})$ be the configuration of a tensegrity cell and $O$ the origin. Let $w_{P_1P_2}$, $w_{P_1P_3}$, ..., $w_{P_4P_5}$ be the self-stress components given by the force-density values assigned to each member. Writing the nodal equilibrium for both Type I and Type II cells gives:

Equilibrium at $P_1$: $\quad w_{P_1P_2}\overrightarrow{P_1P_2} + w_{P_1P_3}\overrightarrow{P_1P_3} + w_{P_1P_4}\overrightarrow{P_1P_4} + w_{P_1P_5}\overrightarrow{P_1P_5} = \vec{0}$ (5)

Equilibrium at $P_2$: $\quad w_{P_2P_1}\overrightarrow{P_2P_1} + w_{P_2P_3}\overrightarrow{P_2P_3} + w_{P_2P_4}\overrightarrow{P_2P_4} + w_{P_2P_5}\overrightarrow{P_2P_5} = \vec{0}$ (6)

Equilibrium at $P_3$: $\quad w_{P_3P_1}\overrightarrow{P_3P_1} + w_{P_3P_2}\overrightarrow{P_3P_2} + w_{P_3P_4}\overrightarrow{P_3P_4} + w_{P_3P_5}\overrightarrow{P_3P_5} = \vec{0}$ (7)

Equilibrium at $P_4$: $\quad w_{P_4P_1}\overrightarrow{P_4P_1} + w_{P_4P_2}\overrightarrow{P_4P_2} + w_{P_4P_3}\overrightarrow{P_4P_3} + w_{P_4P_5}\overrightarrow{P_4P_5} = \vec{0}$ (8)

Equilibrium at $P_5$: $\quad w_{P_5P_1}\overrightarrow{P_5P_1} + w_{P_5P_2}\overrightarrow{P_5P_2} + w_{P_5P_3}\overrightarrow{P_5P_3} + w_{P_5P_4}\overrightarrow{P_5P_4} = \vec{0}$ (9)

The solution of the system given by Equations (5) – (9) can be expressed as a function of one of the self-stress components. Without loss of generality, assume that self-stress component $w_{P_1P_2}$ is known.

Applying the 3D cross product of Equation (5) and vector $\overrightarrow{P_1P_4}$ and the dot product of (5) and $\overrightarrow{P_1P_5}$ allows self-stress component $w_{P_1P_3}$ to be expressed as a function of $w_{P_1P_2}$:

$$((w_{P_1P_2}\overrightarrow{P_1P_2} + w_{P_1P_3}\overrightarrow{P_1P_3} + w_{P_1P_4}\overrightarrow{P_1P_4} + w_{P_1P_5}\overrightarrow{P_1P_5}) \times \overrightarrow{P_1P_4}) \cdot \overrightarrow{P_1P_5} = 0$$

$$\Leftrightarrow w_{P_1P_2}[(\overrightarrow{P_1P_2} \times \overrightarrow{P_1P_4}) \cdot \overrightarrow{P_1P_5}] + w_{P_1P_3}[(\overrightarrow{P_1P_3} \times \overrightarrow{P_1P_4}) \cdot \overrightarrow{P_1P_5}] + w_{P_1P_4}\underbrace{[(\overrightarrow{P_1P_4} \times \overrightarrow{P_1P_4}) \cdot \overrightarrow{P_1P_5}]}_{0} + w_{P_1P_5}\underbrace{[(\overrightarrow{P_1P_5} \times \overrightarrow{P_1P_4}) \cdot \overrightarrow{P_1P_5}]}_{0} = 0$$

$$\Leftrightarrow w_{P_1P_3} = -w_{P_1P_2}\frac{[(\overrightarrow{P_1P_2} \times \overrightarrow{P_1P_4}) \cdot \overrightarrow{P_1P_5}]}{[(\overrightarrow{P_1P_3} \times \overrightarrow{P_1P_4}) \cdot \overrightarrow{P_1P_5}]} \tag{10}$$

Now introducing:

$$f(P_1, P_2, P_3, P_4) = \frac{1}{3!}[(\overrightarrow{P_1P_2} \times \overrightarrow{P_1P_3}) \times \overrightarrow{P_1P_4}]$$

$$= \frac{1}{3!}[((\overrightarrow{P_1O} + \overrightarrow{OP_2}) \times (\overrightarrow{P_1O} + \overrightarrow{OP_3}) \times (\overrightarrow{P_1O} + \overrightarrow{OP_4}))]$$

$$= \frac{1}{3!}[(\overrightarrow{OP_2} \times \overrightarrow{OP_3}) \times \overrightarrow{OP_4}] - [(\overrightarrow{OP_1} \times \overrightarrow{OP_3}) \times \overrightarrow{OP_4}] + [(\overrightarrow{OP_1} \times \overrightarrow{OP_2}) \times \overrightarrow{OP_4}] - [(\overrightarrow{OP_1} \times \overrightarrow{OP_2}) \times \overrightarrow{OP_3}]$$

$$= \frac{1}{3!}\left(\begin{vmatrix} p_{21} & p_{22} & p_{23} \\ p_{31} & p_{32} & p_{33} \\ p_{41} & p_{42} & p_{43} \end{vmatrix} - \begin{vmatrix} p_{11} & p_{12} & p_{13} \\ p_{31} & p_{32} & p_{33} \\ p_{41} & p_{42} & p_{43} \end{vmatrix} + \begin{vmatrix} p_{11} & p_{12} & p_{13} \\ p_{21} & p_{22} & p_{23} \\ p_{41} & p_{42} & p_{43} \end{vmatrix} - \begin{vmatrix} p_{11} & p_{12} & p_{13} \\ p_{21} & p_{22} & p_{23} \\ p_{31} & p_{32} & p_{33} \end{vmatrix}\right)$$

$$= \frac{1}{3!}\begin{vmatrix} 1 & p_{11} & p_{12} & p_{13} \\ 1 & p_{21} & p_{22} & p_{23} \\ 1 & p_{31} & p_{32} & p_{33} \\ 1 & p_{41} & p_{42} & p_{43} \end{vmatrix} \tag{11}$$

the component $w_{P_1P_3}$ can be written as:

$$w_{P_1P_3} = -w_{P_1P_2}\frac{\begin{vmatrix} 1 & p_{11} & p_{12} & p_{13} \\ 1 & p_{21} & p_{22} & p_{23} \\ 1 & p_{41} & p_{42} & p_{43} \\ 1 & p_{51} & p_{52} & p_{53} \end{vmatrix}}{\begin{vmatrix} 1 & p_{11} & p_{12} & p_{13} \\ 1 & p_{31} & p_{32} & p_{33} \\ 1 & p_{41} & p_{42} & p_{43} \\ 1 & p_{51} & p_{52} & p_{53} \end{vmatrix}} = -w_{P_1P_2}\frac{f(P_1, P_2, P_4, P_5)}{f(P_1, P_3, P_4, P_5)} \tag{12}$$

Function $f(P_1,P_2,P_3,P_4)$ reflects the volume of tetrahedron $P_1P_2P_3P_4$. It is thus null if and only if points $P_1,P_2,P_3,P_4$ lie on the same plane or three of them lie on the same line.

Repeating the process for Equations (6) through (9) allows all self-stress components to be expressed as a function of component $w_{P_1P_2}$. Assuming the force-density in element $P_1P_2$ is equal to $\alpha$, the self-stress state $w$ in the cell is given by:

$$w = \begin{bmatrix} w_{P_1P_2} \\ w_{P_2P_3} \\ w_{P_1P_3} \\ w_{P_1P_4} \\ w_{P_2P_4} \\ w_{P_3P_4} \\ w_{P_1P_5} \\ w_{P_2P_5} \\ w_{P_3P_5} \\ w_{P_4P_5} \end{bmatrix} = \alpha \begin{bmatrix} 1 \\ \dfrac{f(P_1,P_2,P_4,P_5)}{f(P_2,P_3,P_4,P_5)} \\ -\dfrac{f(P_1,P_2,P_4,P_5)}{f(P_1,P_3,P_4,P_5)} \\ \dfrac{f(P_1,P_2,P_3,P_5)}{f(P_1,P_3,P_4,P_5)} \\ -\dfrac{f(P_1,P_2,P_3,P_5)}{f(P_2,P_3,P_4,P_5)} \\ \dfrac{f(P_1,P_2,P_3,P_5)}{f(P_1,P_3,P_4,P_5)} \times \dfrac{f(P_1,P_2,P_4,P_5)}{f(P_2,P_3,P_4,P_5)} \\ -\dfrac{f(P_1,P_2,P_3,P_4)}{f(P_1,P_3,P_4,P_5)} \\ \dfrac{f(P_1,P_2,P_3,P_4)}{f(P_2,P_3,P_4,P_5)} \\ -\dfrac{f(P_1,P_2,P_3,P_4)}{f(P_1,P_3,P_4,P_5)} \times \dfrac{f(P_1,P_2,P_4,P_5)}{f(P_2,P_3,P_4,P_5)} \\ \dfrac{f(P_1,P_2,P_3,P_5)}{f(P_1,P_3,P_4,P_5)} \times \dfrac{f(P_1,P_2,P_3,P_4)}{f(P_2,P_3,P_4,P_5)} \end{bmatrix} \quad (13)$$

Equation 13 represents the general expression of the solution for the self-stress in a cell at any configuration in general position. The expression reveals the link between topology and geometry in the self-equilibrium of tensegrity cells and thus the structures they compose, with topology dictating the number of self-stress states and the geometry defining the magnitude in the self-stress components and thus element typology.

## 3. Morphogenesis of tensegrity structures

### 3.1. Cellular analogy

Tensegrity cells can be combined to compose complex tensegrity structures by sharing one or more nodes (and consequently edges). Similar to the interaction between biological cells, if no elements of the cells composing a tensegrity structure are removed, the process corresponds to cellular adhesion (Figure 2a). In cellular adhesion, all cells are stable and can function separately [47]. If elements in the cells composing a tensegrity structure are removed after adhesion occurs, the process corresponds to cellular fusion (Figure 2b). In fusion, cells function together as one entity [48]. The amalgamation of multiple tensegrity cells into a tensegrity structure corresponds to morphogenesis: the biological process that controls the spatial distribution of cells during the development of an organism [49].

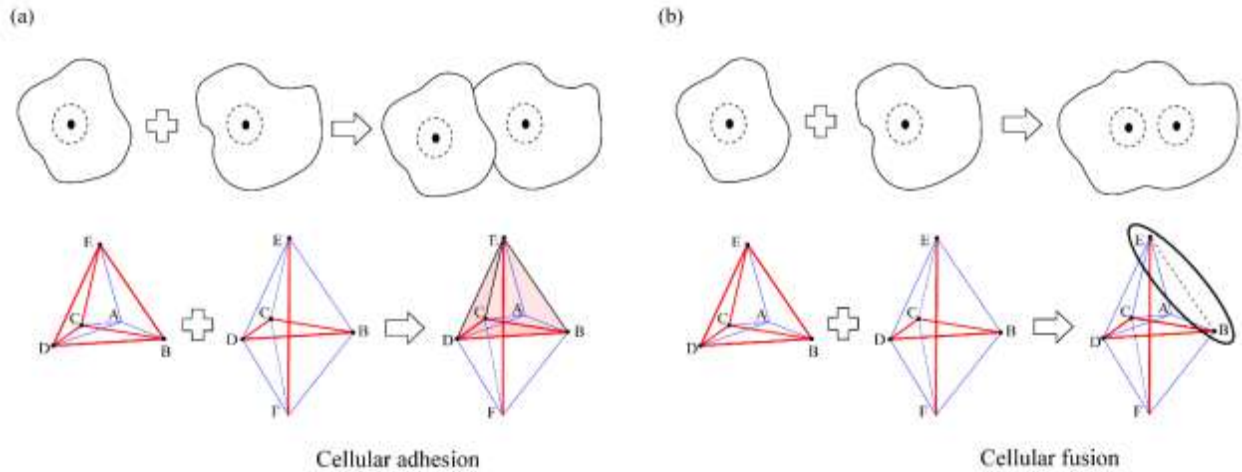

*Figure 2: Illustration of the cellular morphogenesis analogy and its mechanisms.*

3.2. Cellular morphogenesis mechanisms

3.2.1. Cellular adhesion

In the morphogenesis of tensegrity cells, adhesion occurs when two cells are connected and all shared members between the new cell and the existing structure are preserved. Since each cell has five nodes and ten edges, a great number of combinations can occur. Figure 3 illustrates possible combinations for two cells. However, only the cases where the two systems share three or four nodes are of interest, as the other cases either result in finite mechanisms (when the two cells share one or two nodes) or identical structures (when the two cells share all nodes). The adhesion of two cells results always in a rigid structure as the cells are rigid graphs, and adhesion reflects a gluing operation along three or more nodes [50]. Moreover, adhesion increases the number of self-stress states in the structure, as the additional cell can still function independently from the rest of the structure. The self-stress state corresponding to the new cell can be obtained by assigning the self-stress components of elements composing the cell with values obtained from Equation 13 and setting all other self-stress components to zero. Thus, the new state is only a function of the geometry of the new cell. In the case of adhesion, the construction of the self-stress space is simplified and the members' typology depends only on the assignment of the self-stress in the cells.

3.2.2. Cellular fusion

The fusion mechanism occurs when after the connection of two cells, edges are removed. A removed edge can be thought of as a member with zero self-stress. Therefore, the cell being added to the structure should be constructed such that the self-stress coefficients corresponding to the members being removed have opposite signs and are equal in magnitude. The result of fusion mechanism depends on the number of nodes shared between the new cell and the existing structure, as well as the number of removed edges.

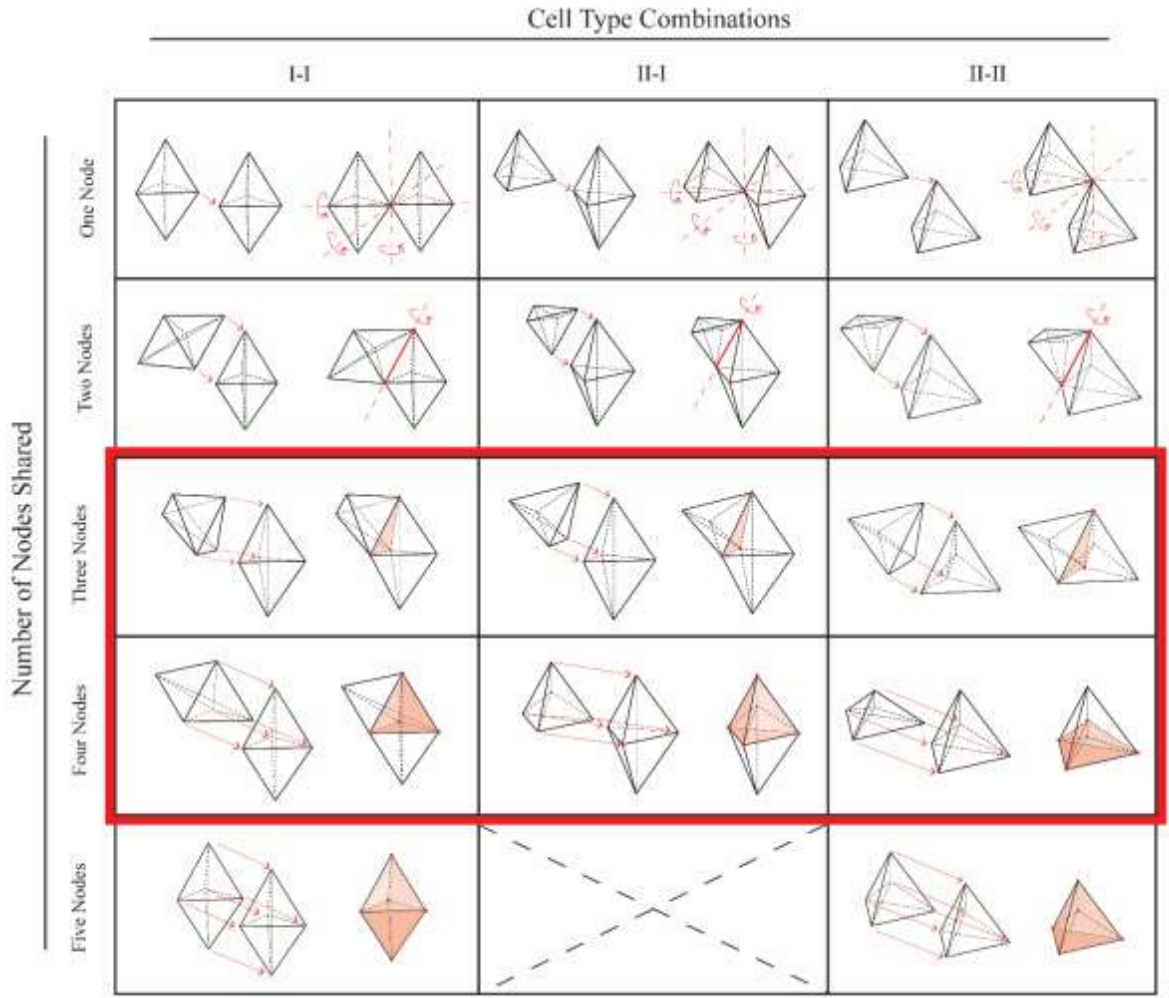

*Figure 3: Illustration of two-cell combinations.*

The removal of an edge is always possible as the resulting structure is always rigid (see Appendix B). The removal can be done by adjusting the self-stress in the new cell (multiplying it with the appropriate scalar coefficient) so that the forces in the elements between the cell and the existing structure cancel out. The nodal coordinates of the cell being added are thus not constrained and can be freely defined.

The removal of multiple edges that belong to the same unicellular tensegrity organisms is only possible when the nodal positions satisfy a set of geometric relations that depend on the geometry of the structure and guarantee the existence of a proper self-stress. However, the stability of the resulting structure should be confirmed. Without loss of generality, consider that cell ABCDE is the new cell being added to an existing structure with shared nodes A, B, C if it shares three nodes with the existing structure or A, B, C, D if it shares four nodes. In the analysis of the removal of two edges, two cases have to be considered: a) if the two edges that are being removed share a node, and b) if the two edges that are being removed do not share a node.

a) If the two edges that are being removed share a node

Let us assume that edges AB and BC are being removed. This implies that the cell ABCDE has to have $w_1$ and $w_2$ for force densities in the member AB and BC, respectively. Thus, the nodal positions of A,B,C,D,E have to satisfy the system:

$$\begin{cases} w_{AB} = -w_1 \\ w_{BC} = w_{AB} \dfrac{f(A,B,D,E)}{f(B,C,D,E)} = -w_2 \end{cases} \Leftrightarrow \begin{cases} w_{AB} = -w_1 \\ w_{BC} = w_{AB} \dfrac{\begin{vmatrix} 1 & a_1 & a_2 & a_3 \\ 1 & b_1 & b_2 & b_3 \\ 1 & d_1 & d_2 & d_3 \\ 1 & e_1 & e_2 & e_3 \end{vmatrix}}{\begin{vmatrix} 1 & b_1 & b_2 & b_3 \\ 1 & c_1 & c_2 & c_3 \\ 1 & d_1 & d_2 & d_3 \\ 1 & e_1 & e_2 & e_3 \end{vmatrix}} = -w_2 \end{cases} \quad (14)$$

Assuming that the cell shares three nodes with the existing structure (i.e. A, B and C), the coordinates of those nodes are already defined since they are part of the existing structure. The coordinates of nodes D and E can be obtained by solving Equation 15. In Equation 15, coordinates $(d_1,d_2,d_3)$ and $(e_1,e_2,e_3)$ were changed to $(x_D,y_D,z_D)$ and $(x_E,y_E,z_E)$ to distinguish between knowns and unknowns.

$$\underbrace{\left(\begin{vmatrix} 1 & a_3 \\ 1 & b_3 \end{vmatrix} - \dfrac{w_2}{w_1}\begin{vmatrix} 1 & b_3 \\ 1 & c_3 \end{vmatrix}\right)}_{\alpha} \begin{vmatrix} x_D & x_E \\ y_D & y_E \end{vmatrix} - \underbrace{\left(\begin{vmatrix} 1 & a_2 \\ 1 & b_2 \end{vmatrix} - \dfrac{w_2}{w_1}\begin{vmatrix} 1 & b_2 \\ 1 & c_2 \end{vmatrix}\right)}_{\beta} \begin{vmatrix} x_D & x_E \\ z_D & z_E \end{vmatrix} + \underbrace{\left(\begin{vmatrix} 1 & a_1 \\ 1 & b_1 \end{vmatrix} - \dfrac{w_2}{w_1}\begin{vmatrix} 1 & b_1 \\ 1 & c_1 \end{vmatrix}\right)}_{\gamma} \begin{vmatrix} y_D & y_E \\ z_D & z_E \end{vmatrix}$$

$$= \underbrace{\left(\begin{vmatrix} a_2 & a_3 \\ b_2 & b_3 \end{vmatrix} - \dfrac{w_2}{w_1}\begin{vmatrix} b_2 & b_3 \\ c_2 & c_3 \end{vmatrix}\right)}_{\omega}(x_E - x_D) - \underbrace{\left(\begin{vmatrix} a_1 & a_3 \\ b_1 & b_3 \end{vmatrix} - \dfrac{w_2}{w_1}\begin{vmatrix} b_1 & b_3 \\ c_1 & c_3 \end{vmatrix}\right)}_{\varphi}(y_E - y_D) + \underbrace{\left(\begin{vmatrix} a_1 & a_2 \\ b_1 & b_2 \end{vmatrix} - \dfrac{w_2}{w_1}\begin{vmatrix} b_1 & b_2 \\ c_1 & c_2 \end{vmatrix}\right)}_{\psi}(z_E - z_D)$$

$$\Leftrightarrow \alpha \begin{vmatrix} x_D & x_E \\ y_D & y_E \end{vmatrix} - \beta \begin{vmatrix} x_D & x_E \\ z_D & z_E \end{vmatrix} + \gamma \begin{vmatrix} y_D & y_E \\ z_D & z_E \end{vmatrix} = \omega(x_E - x_D) - \varphi(y_E - y_D) + \psi(z_E - z_D)$$

$$\Leftrightarrow \begin{bmatrix} x_D & y_D & z_D & 1 \end{bmatrix} \begin{bmatrix} 0 & \alpha & \beta & \omega \\ -\alpha & 0 & \gamma & \varphi \\ -\beta & -\gamma & 0 & \psi \\ -\omega & -\varphi & -\psi & 0 \end{bmatrix} \begin{bmatrix} x_E \\ y_E \\ z_E \\ 1 \end{bmatrix} = 0 \quad (15)$$

On the other hand, if the cell shares four nodes with the existing structure (assuming that node D is also part of the existing structure along with nodes A,B, and C the coordinates of node E can be defined by:

$$\begin{bmatrix} d_1 & d_2 & d_3 & 1 \end{bmatrix} \begin{bmatrix} 0 & \alpha & \beta & \omega \\ -\alpha & 0 & \gamma & \varphi \\ -\beta & -\gamma & 0 & \psi \\ -\omega & -\varphi & -\psi & 0 \end{bmatrix} \begin{bmatrix} x_E \\ y_E \\ z_E \\ 1 \end{bmatrix} = 0 \quad (16)$$

Equation 16 represents a planar surface with any point belonging to this plane being a valid solution for the position of node E.

b) If the two edges that are being removed do not share a node

When the cell being added and the existing structure share four nodes, and the two edges that are being removed do not share a node (i.e. AB and CD), the position of node E must satisfy the system:

$$\begin{cases} w_{AB} = -w_1 \\ w_{CD} = w_{AB} \dfrac{f(A,B,C,E)}{f(A,C,D,E)} \times \dfrac{f(A,B,D,E)}{f(B,C,D,E)} = -w_2 \end{cases} \Leftrightarrow \begin{cases} w_{AB} = -w_1 \\ w_{CD} = w_{AB} \dfrac{\begin{vmatrix} 1 & a_1 & a_2 & a_3 \\ 1 & b_1 & b_2 & b_3 \\ 1 & c_1 & c_2 & c_3 \\ 1 & e_1 & e_2 & e_3 \end{vmatrix} \begin{vmatrix} 1 & a_1 & a_2 & a_3 \\ 1 & b_1 & b_2 & b_3 \\ 1 & d_1 & d_2 & d_3 \\ 1 & e_1 & e_2 & e_3 \end{vmatrix}}{\begin{vmatrix} 1 & a_1 & a_2 & a_3 \\ 1 & c_1 & c_2 & c_3 \\ 1 & d_1 & d_2 & d_3 \\ 1 & e_1 & e_2 & e_3 \end{vmatrix} \begin{vmatrix} 1 & b_1 & b_2 & b_3 \\ 1 & c_1 & c_2 & c_3 \\ 1 & d_1 & d_2 & d_3 \\ 1 & e_1 & e_2 & e_3 \end{vmatrix}} = -w_2 \end{cases} \quad (17)$$

$F(A,B,C,D)$ denotes the matrix whose determinant is defined in Equation (11) and $\Delta_{ij}^{ABCD}$ is the cofactor of the matrix $F(A,B,C,D)$ defined by:

$$\Delta_{ij}^{ABCD} = (-1)^{i+j} \left| M_{ij}^{ABCD} \right|$$

$M_{ij}^{ABCD}$ is the minor of $F(A,B,C,D)$ obtained by deleting row $i$ and column $j$. Node E belongs thus to the quadratic surface defined by:

$$\begin{bmatrix} 1 & x_E & y_E & z_E \end{bmatrix} T \begin{bmatrix} 1 \\ x_E \\ y_E \\ z_E \end{bmatrix} = 0 \quad (18)$$

where the components of the matrix $T$ are (see Appendix A for details):

$$T_{ij} = \Delta_{4i}^{ABCE} \times \Delta_{4j}^{ABDE} - \frac{w_2}{w_1} \Delta_{4i}^{BCDE} \times \Delta_{4j}^{ACDE} \quad (19)$$

Since the expressions for the self-stress components (Equation 13) do not depend on node labeling, any permutation of the node labels results in expressions of the same form. Moreover, the solution space for the added nodes can be found through the intersection of the surfaces defined by considering removing all the combinations of two edges. Consequently, if three edges need to be removed, three subspaces can be defined by considering every combination of two edges and the added nodes will be on the intersection of these spaces.

### 3.3. Morphogenesis and self-stress

Cellular morphogenesis reflects the reverse process of the tensegrity decomposition proposed by de Guzmán and Orden 2006 [6]. A corollary to a Proposition by Fernández and Orden (2011) [51], which allows one to combinatorially calculate the number of self-stress states by decomposing tensegrity structures into cells, was proposed by Aloui et al. (2018) [30] and is adapted here for three-dimensional tensegrity structures:

**Corollary.** Let $G_i$ and $G_{i+1}$ be the abstract underlying graphs of the tensegrity structures obtained through cellular morphogenesis at steps $i$ and $i+1$. Let $B_i$ and $B_{i+1}$ be their Laman bounds, respectively. Let $e_i$ be the change in the number of edges between $G_i$ and $G_{i+1}$, and $v_i$ be the change in the number of nodes. Assuming that $G_{i+1}$ is generically rigid, the change in the dimension of the self-stress space $W$ is given by:

$$\Delta(\dim(W)) = e_i - 3v_i$$

The corollary can be used to identify changes in the number of self-stress states during adhesion and fusion. For three-dimensional tensegrity structures, the number of added or removed states can take any integer value from 0 to 10 depending on the change in the number of edges and nodes of the structure, with each state reflecting the contribution of a single unicellular organism. These unicellular organisms can be tensegrity cells if they are complete graphs on five nodes or virtual cells [30]: subgraphs with one self-stress state formed by the interactions between cells. For cells, the self-stress state can be calculated through Equation 13. However, the self-stress state corresponding to virtual cells has to be calculated through the nullspace of the equilibrium matrix as, contrary to the planar case, a specific pattern cannot be identified. A cellular morphogenesis example is presented below to elucidate the construction of a basis for the self-stress space and the effects of adhesion and fusion in the space.

Morphogenesis starts with a Type I cell, denoted as Cell 1, defined by the nodes {1,2,3,4,5}. The self-stress in Cell 1 is described by the vector $w_1$. In step one, a second Type I cell {2,3,4,5,6}, named Cell 2, is added to the existing cell with the two cells sharing four nodes. The adhesion of the second cell results in a second self-stress state $w_2$ (second column of the matrix $W$). Each state (column of matrix $W$) corresponds to the self-stress coefficients that stabilize the corresponding cell with zeros for edges that are not part of the cell. In step two, the adhesion of a third Type I cell {1,2,3,7,6}, named Cell 3, results into two additional self-stress states $w_3$ and $w_4$: the first state (third column of matrix $W$) stabilizing Cell 3 itself, while the second (fourth column of matrix $W$) reflects the presence of a virtual cell {1,2,3,4,6} (Cell 4).

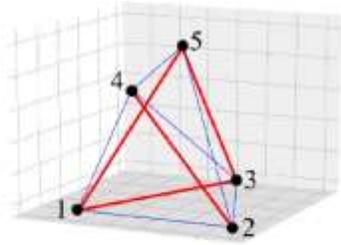
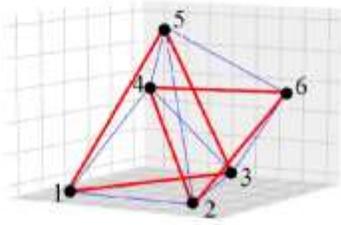
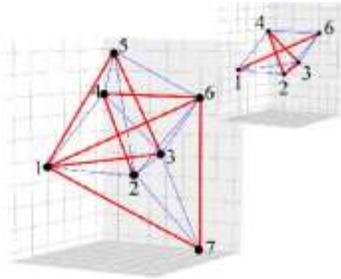

*Figure 4: Illustration of the construction of a self-stress basis in the case of adhesion steps.*

Assume that fusion occurs in the third step with the removal of edge (2,3), which is shared by all four unicellular organisms. In order to remove edge (2,3), all combinations of Cell 3 and the other cells must be considered, leading to three self-stress states as shown in Figure 5. The result of the fusion is a structure composed of three unicellular organisms with each one being the result of the fusion of two cells. The fusion of Cells 1 and 3 forms the first unicellular organism stabilized by the self-stress state described by the linear combination $\left(w_1 + \frac{1.635}{1.25} w_3\right)$ allowing the cancelation of the self-stress component of element (2,3) and thus its removal. Similarly, the fusion of Cells 2 and 3, and the fusions of Cells 3 and 4 result in the self-stress states $\left(w_2 + \frac{1}{1.25} w_3\right)$ and $\left(w_4 - \frac{0.247}{1.25} w_3\right)$ which allow the removal of element (2,3).

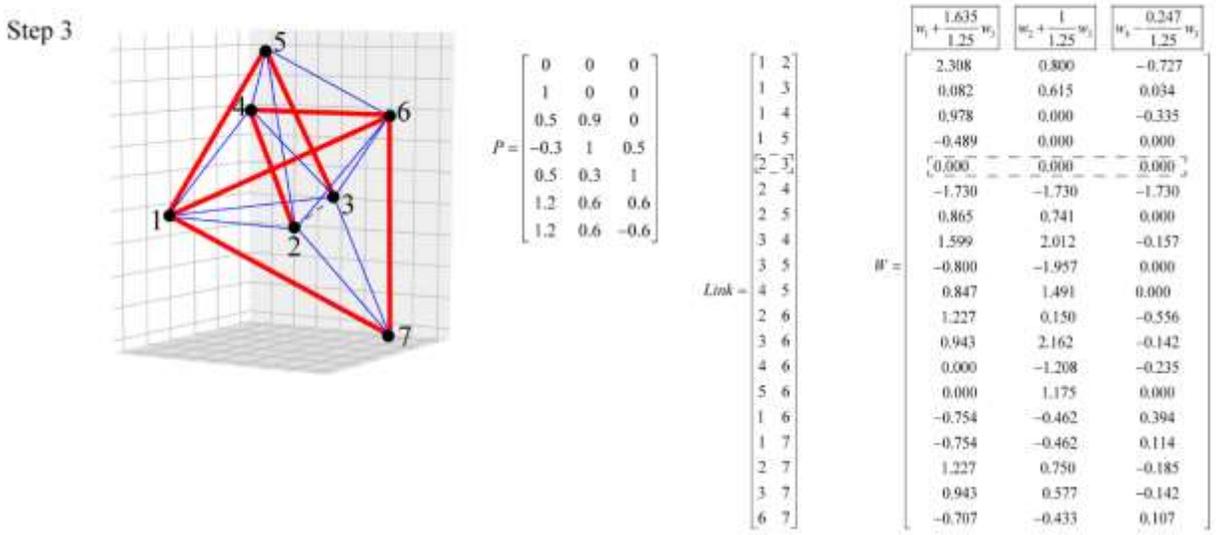

*Figure 5: Illustration of the construction of self-stress basis in the case of a fusion step.*

The example reveals that the number of unicellular organisms (cells being added and virtual cells) combined with the number of adhesion and fusion steps define the number of self-stress states, while their magnitudes depend on their geometry.

## 4. Implementation of the Cellular Morphogenesis for Tensegrity Structures

Cellular morphogenesis of tensegrity structures is implemented exclusively using graphs, as both the tensegrity structures and the process itself are modeled as simple undirected graphs. Tensegrity structures are modeled using a graph $G(V,E)$ where the set of vertices $V$ represents the nodes of the structure and the set of edges E describe the members of the structure. The morphogenesis process is modeled as a graph $G_c(V_c, E_c)$ where $V_c$ is the set of the cells and unicellular organisms employed during the generative process and $E_c$ is the set of edges representing the shared boundary between cells (Figure 6).

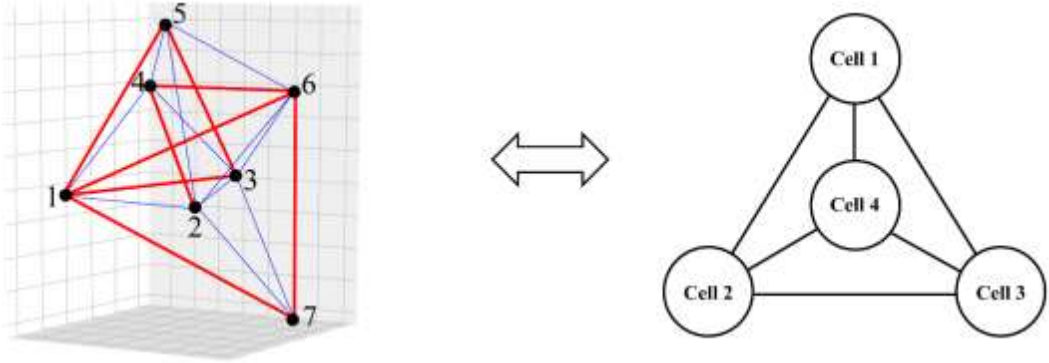

*Figure 6: Graph Gc of the cellular morphogenesis (vertices are cells, edges are shared boundary).*

Nodes in graph $G_c$ correspond to the cells composing the tensegrity structure generated. Therefore, the nodal attributes (Table 1) include a cell type ("regular cell" if it was an added cell, "virtual cell" if it corresponds to a unicellular organism identified during adhesion for the completion of the self-stress basis, or "fused cell" if it is a unicellular structure created after the fusion of two cells), as well as the nodes and connectivity for each cell. Edge attributes (Table 2) include the set of shared members between cells *i* and *j*. Employing graph models for the implementation of the method is advantageous not only because of the nature of the structures being modeled, but also because the construction of the self-stress space and the identification of the virtual cells patterns depend on the history of the morphogenesis steps (cells being formed and transformed throughout the process). The cellular morphogenesis method was thus implemented using Python [52] and Networkx [53], a Python package for graph theory.

*Table 1: Nodal attributes of the graph $G_c$.*

| Cell No. | Type | Cell Graph | |
|---|---|---|---|
| | | Nodes | Edges |
| Cell 1 | Regular | {1,2,3,4,5} | {(1,2), (1,3), (1,4), (1,5), (2,3), (2,4), (2,5), (3,4), (3,5), (4,5)} |
| Cell 2 | Regular | {2,3,4,5,6} | {(2,3), (2,4), (2,5), (2,6), (3,4), (3,5), (3,6), (4,5), (4,6), (5,6)} |
| Cell 3 | Regular | {1,2,3,6,7} | {(1,2), (1,3), (1,6), (1,7), (2,3), (2,6), (2,7), (3,6), (3,7), (6,7)} |
| Cell 4 | Virtual | {1,2,3,4,6} | {(1,2), (1,3), (1,4), (1,6), (2,3), (2,4), (2,6), (3,4), (3,6), (4,6)} |

*Table 2: Edges attributes of the graph $G_c$.*

| Edge Id | Shared members |
|---|---|
| (Cell 1, Cell 2) | (2,3), (3,4), (4,5), (2,5), (2,4), (3,5) |
| (Cell 1, Cell 3) | (1,2), (1,3), (2,3) |
| (Cell 1, Cell 4) | (1,2), (2,3), (3,4), (1,4), (2,4), (1,3) |
| (Cell 2, Cell 3) | (2,3), (3,6), (2,6) |
| (Cell 2, Cell 4) | (2,3), (3,4), (4,6), (2,6), (2,4), (3,6) |
| (Cell 3, Cell 4) | (1,2), (1,3), (2,3) |

The flow chart in Figure 7 describes the cellular morphogenesis process of tensegrity structures. Starting from a cell, the input for the cell to be added is provided: the set of shared nodes with the existing structure and the coordinates of the new nodes. An adhesion routine is then called. In the adhesion routine, using the corollary, the existence of virtual cell(s) is determined: if the change in the dimension of the self-stress space is larger than one, then one or multiple virtual cells may be present. The graph $G$ corresponding to the tensegrity structure being generated and the graph $G_c$ corresponding to the cells formed during the process are then updated along with the self-stress. If edges are to be removed from the structure, a fusion routine is called. The fusion routine takes as input the structure and the edges to be removed. In the case of removing more than one edge, the fusion routine determines the geometric relations that the new nodes should satisfy and updates their positions accordingly. Similarly to the adhesion routine, the graph $G$ corresponding to the tensegrity structure being generated and the graph $G_c$ corresponding to the cells formed during the process are updated along with the self-stress. The process is repeated until the desired tensegrity structure is obtained.

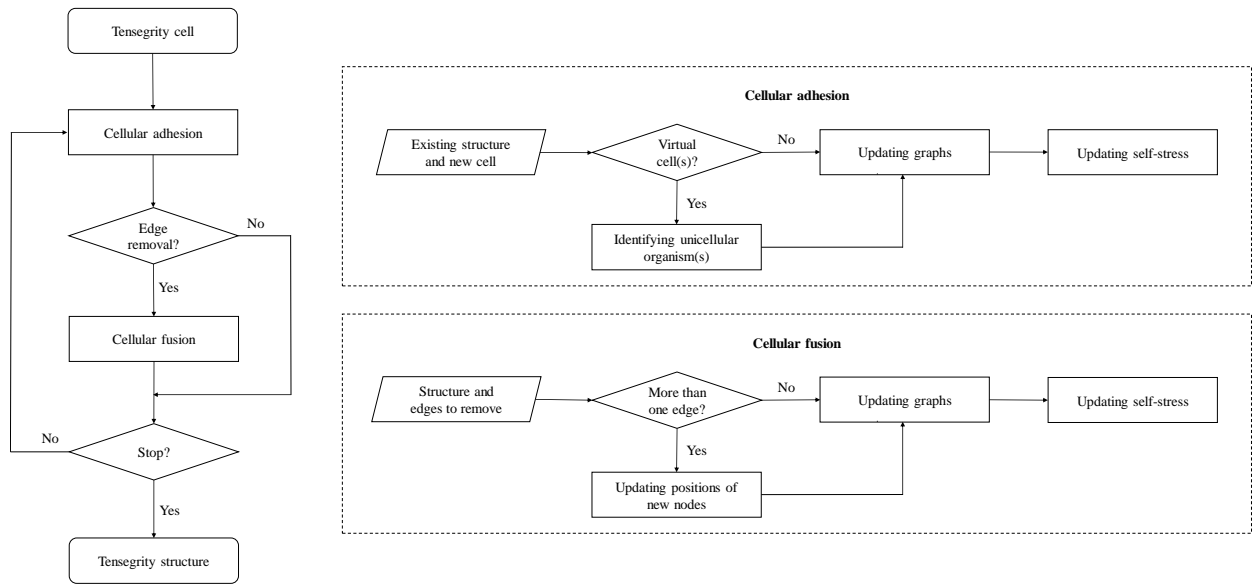

*Figure 7: Flow chart of the cellular morphogenesis process.*

The update of the self-stress is based on the number of cells and virtual cells formed at each step. For cells, the self-stress state is calculated through Equation 13, while for virtual cells the self-stress state is calculated through the nullspace of their equilibrium. Therefore, a routine that extracts subgraphs from the total graph capable of underlying a structure with one self-stress state is proposed. The idea behind the routine is that the self-stress corresponding to the substructure must complete the basis describing the self-stress space in the structure (it must be linearly independent with the set of existing self-stress vectors). The search routine for virtual cells in three-dimensional structures is presented below.

Let $s$ be the number of self-stress states at a step $i$ and $p$ the number of regular cells employed in the generation of a structure until step $i$.

i. Remove an edge belonging to one cell only from each regular or fused cell composing the structure. If the edge removed is attached to a node of degree four (with four elements connected to it), the removal of the edge can result in the removal of the node and thus the removal of the other three edges attached to it. The removal of an edge, or four edges and a node, always results in decreasing the dimension of the self-stress space by one. The number of self-stress states remaining in the structure is *s-p*.
ii. From the remaining structure, select *s-p-1* edges to remove. Preferably, start with edges that have both end nodes with a degree equal or larger than five. The remaining structure will be a unicellular organism with exactly one self-stress state.
iii. Calculate the self-stress state in the unicellular organism by finding the nullspace of its equilibrium matrix. Check if the state is linearly independent with the existing states, if not discard the structure. If it is linearly independent, update the self-stress space of the structure.
iv. Go back to step (ii) and repeat the process by considering a different set of edges until the required number of virtual cells dictated by the corollary has been identified.

## 5. Examples of tensegrity structures generated with cellular morphogenesis

### 5.1. Triplex

The Triplex (also known as Simplex or Tensegrity prism) is an elementary three-dimensional tensegrity structure with six nodes, nine elements in tension, and three elements in compression. The Triplex can be obtained through form finding of a straight triangular prism with its stable equilibrium configuration being when the base triangles belong to parallel planes and they have an angle of twist of π/6 (Figure 8).

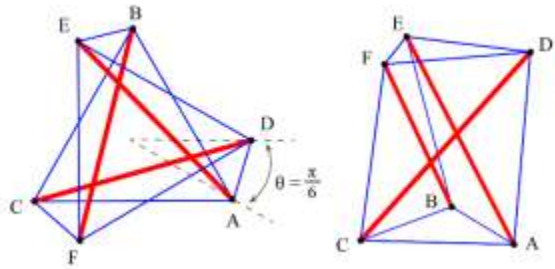

*Figure 8: The Triplex structure in its equilibrium configuration.*

The Triplex has one self-stress state given by:

$$w = w_1[-\sqrt{3},-\sqrt{3},-\sqrt{3},\sqrt{3},\sqrt{3},\sqrt{3},1,1,1,1,1,1]^T \tag{20}$$

where $w_1$ is the self-stress in a base element (members are ordered such that the first three elements are struts, followed by the three lateral cables and the six horizontal cables).

In this study, the Triplex is obtained through cellular morphogenesis. The structure can be obtained through the adhesion of two cells that share four nodes and their fusion with the removal of two edges. Figure 9 illustrates the composition of a Triplex starting from two Type I cells ABCDE and BCDEF. The two cells are combined together by sharing nodes BCDE and then elements BD and CE are removed.

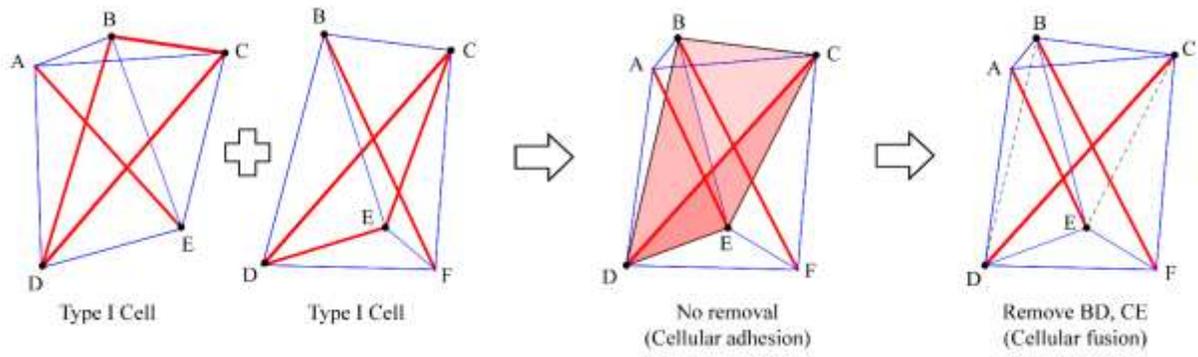

*Figure 9: Cellular morphogenesis of the Triplex.*

The construction of a basis for the self-stress state of a Triplex is presented in Table 3.

*Table 3: Construction of the self-stress state of a Triplex.*

| Cell ABCDE | | Cell BCDEF | | Adhesion step | | | Fusion step | |
|---|---|---|---|---|---|---|---|---|
| Member | Self-stress | Member | Self-stress | Member | Self-stress I | Self-stress II | Member | Self-stress |
| AB | 1.155 | BC | 1.732 | AB | 1.155 | 0.000 | AB | 1.155 |
| BC | -0.577 | CD | -1.000 | BC | -0.577 | 1.732 | BC | 1.155 |
| AC | 1.155 | BD | 1.000 | AC | 1.155 | 0.000 | AC | 1.155 |
| AD | 2.000 | BE | 1.000 | AD | 2.000 | 0.000 | AD | 2.000 |
| BD | -1.000 | CE | -1.000 | BD | -1.000 | 1.000 | BD | 0.000 |
| CD | -1.000 | DE | -0.577 | CD | -1.000 | -1.000 | CD | -2.000 |
| AE | -2.000 | BF | -2.000 | AE | -2.000 | 0.000 | AE | -2.000 |
| BE | 1.000 | CF | 2.000 | BE | 1.000 | 1.000 | BE | 2.000 |
| CE | 1.000 | DF | 1.155 | CE | 1.000 | -1.000 | CE | 0.000 |
| DE | 1.732 | EF | 1.155 | DE | 1.732 | -0.577 | DE | 1.155 |
| | | | | BF | 0.000 | -2.000 | BF | -2.000 |
| | | | | CF | 0.000 | 2.000 | CF | 2.000 |
| | | | | DF | 0.000 | 1.155 | DF | 1.155 |
| | | | | EF | 0.000 | 1.155 | EF | 1.155 |

Following the principles of cellular morphogenesis, the structure ABCDEF resulting from the adhesion of the two cells results into two self-stress states (Table 3, adhesion step). The removal of edges BD and CE (fusion) decreases the number of self-stress states to one. The resulting self-stress state (Table 3, fusion step) is collinear to the vector described in Equation 20 (after setting the same order of elements). In the triplex case, the two removed edges do not share a node. This falls into the case where the position of the added node lies on a quadratic surface defined by Equation 18. When the nodal positions of the Triplex are used, the resulting curve is given by:

$$z^2 - 2xy + xz + yz - z - y = 0 \qquad (21)$$

Figure 10 illustrates the equilibrium surface with the Triplex and all its nodes lying on the meshed surface given by Equation 21.

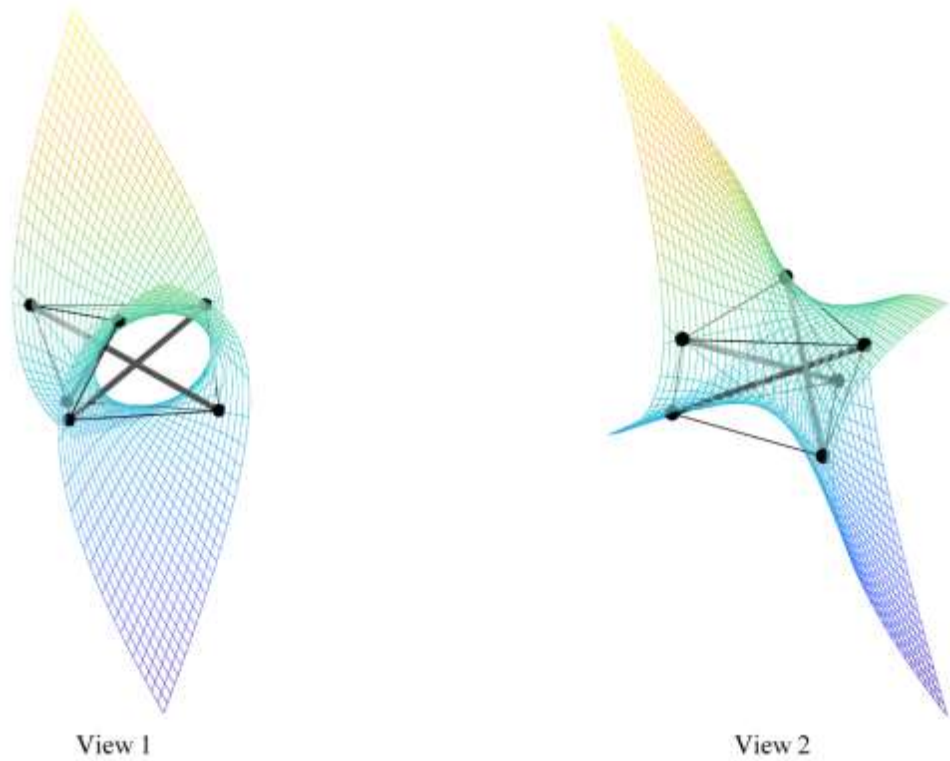

*Figure 10: Illustration of the nodal equilibrium geometric conditions of a regular Triplex.*

### 5.2. Icosahedron

The Icosahedron (also known as the expanded octahedron) is another well-known tensegrity structure. The icosahedron possesses a spherical symmetry which makes it convenient for robotic applications such as NASA's SuperBall developed for planetary landing and exploration [12,54]. The Icosahedron is composed of six struts and twenty-four cables that connect twelve nodes. It has a five-regular graph topology where four cables and one strut are incident to each node. Figure 11 illustrates the cells that compose the Icosahedron along with the related morphogenesis mechanisms and the edges being removed (dashed lines in the central figure). In total, sixteen cells are combined using fifteen adhesion and nine fusion steps, resulting in a structure with one self-stress state (Table 4).

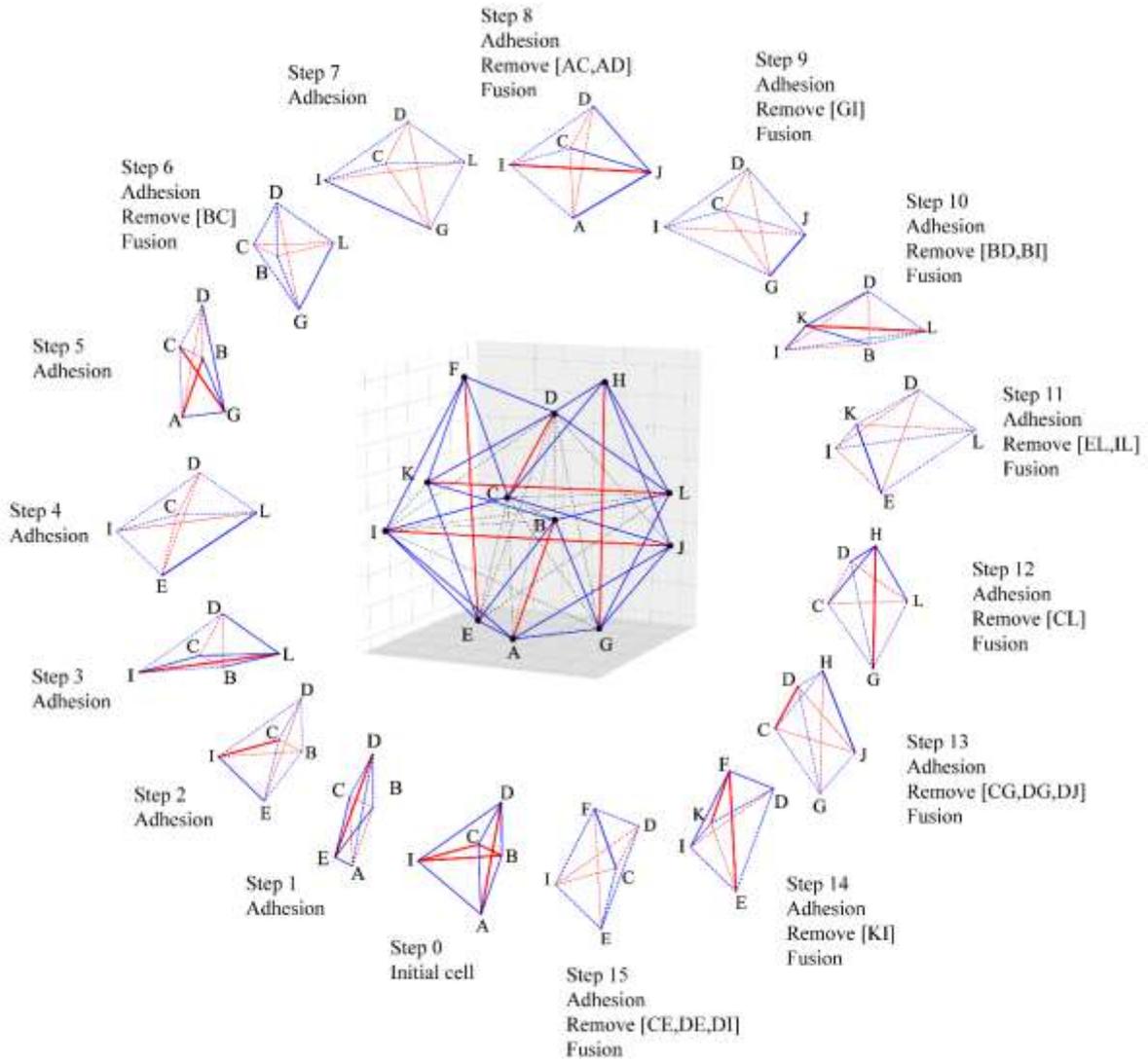

*Figure 11: Illustration of the cellular morphogenesis of the Icosahedron. The dashed elements of the cells are already present in the structure at that step.*

*Table 4: Self-stress state of the Icosahedron.*

| Member | Self-stress coefficient | Member | Self-stress coefficient | Member | Self-stress coefficient |
|---|---|---|---|---|---|
| AB | -1.5 | BE | 1 | DK | 1 |
| CD | -1.5 | BG | 1 | DL | 1 |
| EF | -1.5 | BK | 1 | EI | 1 |
| GH | -1.5 | BL | 1 | EK | 1 |
| IJ | -1.5 | CF | 1 | FI | 1 |
| KL | -1.5 | CH | 1 | FK | 1 |
| AE | 1 | CI | 1 | GJ | 1 |
| AG | 1 | CJ | 1 | GL | 1 |
| AI | 1 | DF | 1 | HJ | 1 |
| AJ | 1 | DH | 1 | HL | 1 |

5.3. Stanford bunny

The Stanford bunny (Figure 12) was developed by Greg Turk and Mark Levoy as a benchmark computer graphics 3D model [55]. In this study, the Stanford bunny was chosen to elucidate the capability of the cellular morphogenesis method in generating complex tensegrity structures with irregular forms.

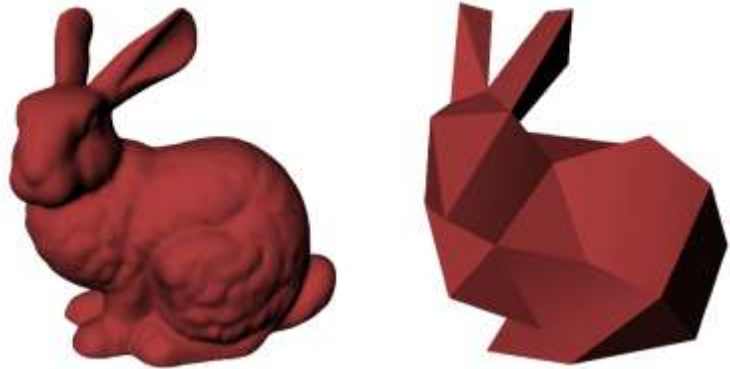

*Figure 12: Illustration of the Stanford bunny structure (left) and the low polygonal surface mesh (right).*

A low cell-resolution bunny was created using a polygonization algorithm. Cells were then created using the 34 nodes of the mesh. The process resulted in 20 Type I cells with 34 nodes and 134 elements (80 elements in tension and 54 elements in compression). The resulting tensegrity structure (Figure 13) has 41 self-stress states with 21 corresponding to virtual cells. A high cell-resolution bunny generated using 528 nodes is presented in Appendix C.

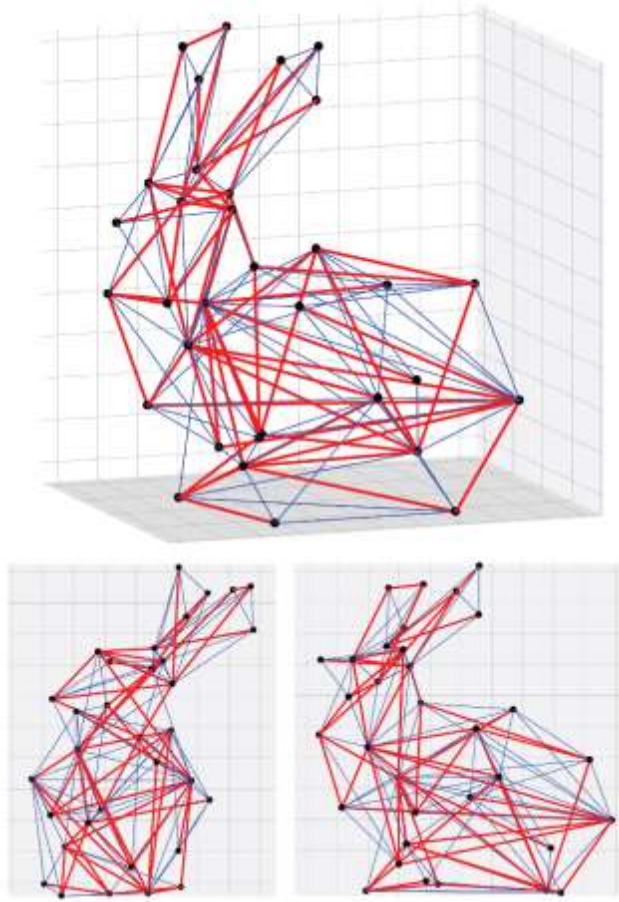

*Figure 13: Illustration of the low-resolution Stanford bunny tensegrity structure.*

A Class 1 bunny with no connection between compression elements (Figure 14) was also constructed using cellular morphogenesis. Although the number of elements in the Class 1 bunny structure is lower than the number of elements in the low cell-resolution bunny, the number of cells employed in the construction of the tensegrity systems is considerably larger as a large number of cells is accounted for the fusion steps required to isolate the compressive elements. The total number of cells used in the morphogenesis process was thus 82 with 61 fusion steps being performed. The resulting bunny structure has 34 nodes and 112 elements (20 in compression and 92 in tension). The resulting structure has 3 self-stress states and 11 infinitesimal mechanisms.

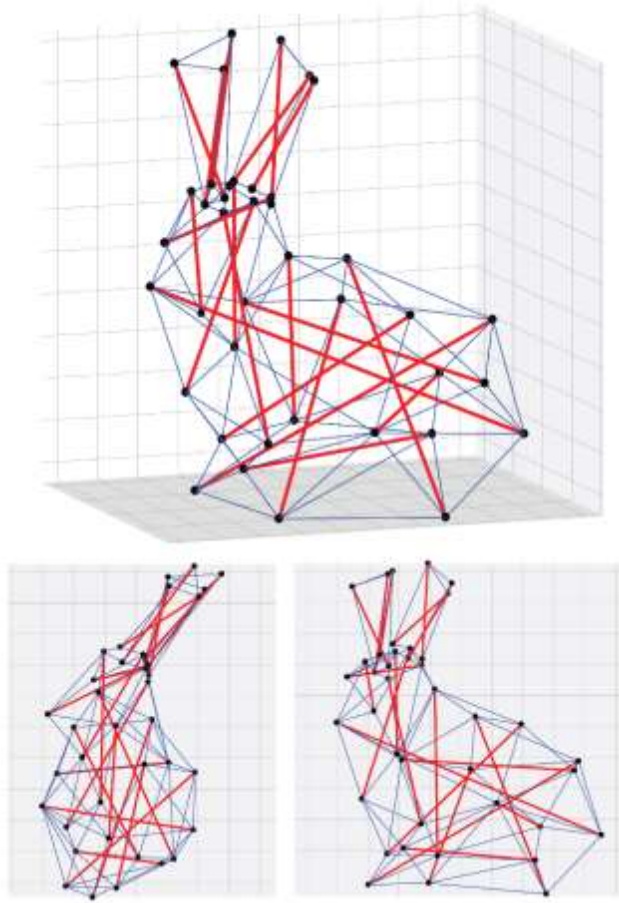

*Figure 14: Illustration of the Class-1 Stanford bunny tensegrity structure.*

## 6. Discussion

Cellular morphogenesis reflects a generative design scheme for complex tensegrity structures based on stable tensegrity cells. Through the cell definition, the method decodes the relation between the topology and geometry in tensegrity structures, allowing for the control of the self-stress states. This can lead to an enhanced design of tensegrity structures with predefined self-stress features. Furthermore, the form of the self-stress states obtained through the cellular morphogenesis conveys a more profound message to the designer than the simple rank analysis of the equilibrium matrix, as each state corresponds to a specific stable unicellular sub-structure. Moreover, through the large number of possible combinations between cells, the proposed method allows for a better exploration of the design space, while also providing a sequence for the assembly and disassembly of the structure. Finally, although cellular morphogenesis of tensegrity structures shows great potential in finding irregular large tensegrity structures with predefined shapes and self-stress states, a deeper understanding of the effects of degeneracy is needed.

## 7. Conclusions

Cellular morphogenesis is novel biomimetic method for the generative design of tensegrity structures that combines topology search and form finding. The method is inspired by the adhesion and fusion of biological cells. The mechanisms of adhesion and fusion are applied to tensegrity cells: infinitesimally rigid tensegrity units of one self-stress state that can compose any tensegrity structure. Through the analytical description of the self-stress in the cells and the study of the impacts of adhesion and fusion on the self-equilibrium of a tensegrity structures, it is shown that topology reflected by the number of unicellular organisms (regular and virtual cells) dictates the number of self-stress states, while the geometry given by their nodal positions defines the magnitude of the forces in those states and thus also element typology. Consequently, cellular morphogenesis offers a new paradigm in the topology search and form finding of tensegrity structures, allowing for the control of the equilibrium geometry and the self-equilibrium through the variation of the adhesion and fusion steps. The method also provides a description of the self-stress space through the construction of a base with vectors in a form that conveys a broader message than a simple rank analysis of the equilibrium matrix. These features of the method can enhance the applications of tensegrity structures in science and engineering.

## 8. Acknowledgements

This material is based upon work supported partially by the National Science Foundation under grant no. 1638336. David Orden has been partially supported by project MTM2017-83750-P of the Spanish Ministry of Science (AEI/FEDER, UE) and by H2020-MSCA-RISE project 734922 - CONNECT.

**Appendix A:**

When the cell being added and the existing structure share four nodes and the two edges that are being removed do not share a node (i.e. AB and CD), the position of node E must satisfy the system:

$$\begin{cases} w_{AB} = -w_1 \\ w_{CD} = w_{AB} \dfrac{f(A,B,C,E)}{f(A,C,D,E)} \times \dfrac{f(A,B,D,E)}{f(B,C,D,E)} = -w_2 \end{cases} \Leftrightarrow \begin{cases} w_{AB} = -w_1 \\ w_{CD} = w_{AB} \dfrac{\begin{vmatrix} 1 & a_1 & a_2 & a_3 \\ 1 & b_1 & b_2 & b_3 \\ 1 & c_1 & c_2 & c_3 \\ 1 & e_1 & e_2 & e_3 \end{vmatrix} \begin{vmatrix} 1 & a_1 & a_2 & a_3 \\ 1 & b_1 & b_2 & b_3 \\ 1 & d_1 & d_2 & d_3 \\ 1 & e_1 & e_2 & e_3 \end{vmatrix}}{\begin{vmatrix} 1 & a_1 & a_2 & a_3 \\ 1 & c_1 & c_2 & c_3 \\ 1 & d_1 & d_2 & d_3 \\ 1 & e_1 & e_2 & e_3 \end{vmatrix} \begin{vmatrix} 1 & b_1 & b_2 & b_3 \\ 1 & c_1 & c_2 & c_3 \\ 1 & d_1 & d_2 & d_3 \\ 1 & e_1 & e_2 & e_3 \end{vmatrix}} = -w_2 \end{cases} \quad (A.1)$$

$F(A,B,C,D)$ denotes the matrix whose determinant is defined in Equation (11) and $\Delta_{ij}^{ABCD}$ is the cofactor of the matrix $F(A,B,C,D)$ defined by:

$$\Delta_{ij}^{ABCD} = (-1)^{i+j} \left| M_{ij}^{ABCD} \right| \tag{A.2a}$$

$M_{ij}^{ABCD}$ is the minor of $F(A,B,C,D)$ obtained by deleting row $i$ and column $j$. Let $(x_1, x_2, x_3)$ be the coordinates of the node E:

$$\begin{vmatrix} 1 & a_1 & a_2 & a_3 \\ 1 & b_1 & b_2 & b_3 \\ 1 & c_1 & c_2 & c_3 \\ 1 & x_1 & x_2 & x_3 \end{vmatrix} = \sum_{i=1}^{4} \Delta_{i4}^{ABCE} x_i \tag{A.2b}$$

System (A.1) becomes:

$$\begin{vmatrix} 1 & a_1 & a_2 & a_3 \\ 1 & b_1 & b_2 & b_3 \\ 1 & c_1 & c_2 & c_3 \\ 1 & x_1 & x_2 & x_3 \end{vmatrix} \begin{vmatrix} 1 & a_1 & a_2 & a_3 \\ 1 & b_1 & b_2 & b_3 \\ 1 & d_1 & d_2 & d_3 \\ 1 & x_1 & x_2 & x_3 \end{vmatrix} = \frac{w_2}{w_1} \begin{vmatrix} 1 & a_1 & a_2 & a_3 \\ 1 & c_1 & c_2 & c_3 \\ 1 & d_1 & d_2 & d_3 \\ 1 & x_1 & x_2 & x_3 \end{vmatrix} \begin{vmatrix} 1 & b_1 & b_2 & b_3 \\ 1 & c_1 & c_2 & c_3 \\ 1 & d_1 & d_2 & d_3 \\ 1 & x_1 & x_2 & x_3 \end{vmatrix}$$

$$\Leftrightarrow \left( \sum_{i=1}^{4} \Delta_{i4}^{ABCE} x_i \right) \left( \sum_{i=1}^{4} \Delta_{i4}^{ABDE} x_i \right) = \frac{w_2}{w_1} \left( \sum_{i=1}^{4} \Delta_{i4}^{ACDE} x_i \right) \left( \sum_{i=1}^{4} \Delta_{i4}^{BCDE} x_i \right)$$

$$\Leftrightarrow \sum_{i=1}^{4} \sum_{j=1}^{4} \Delta_{i4}^{ABCE} \Delta_{j4}^{ABDE} x_i x_j = \frac{w_2}{w_1} \sum_{i=1}^{4} \sum_{j=1}^{4} \Delta_{i4}^{ACDE} \Delta_{j4}^{BCDE} x_i x_j$$

$$\Leftrightarrow \sum_{i=1}^{4} \sum_{j=1}^{4} \left( \Delta_{i4}^{ABCE} \Delta_{j4}^{ABDE} - \frac{w_2}{w_1} \Delta_{i4}^{ACDE} \Delta_{j4}^{BCDE} \right) x_i x_j = 0$$

$$[1 \quad x_E \quad y_E \quad z_E] T \begin{bmatrix} 1 \\ x_E \\ y_E \\ z_E \end{bmatrix} = 0 \tag{A.2c}$$

with

$$T_{ij} = \Delta_{4i}^{ABCE} \times \Delta_{4j}^{ABDE} - \frac{w_2}{w_1} \Delta_{4i}^{BCDE} \times \Delta_{4j}^{ACDE} \tag{A.2d}$$

**Appendix B:**

To prove that the fusion process with the removal of an edge preserves the rigidity of a given tensegrity structure, two results of rigidity theory and graph theory must be considered. The first result is a theorem by Roth and Whiteley (1981) relating the rigidity of tensegrity frameworks to the rigidity of the underlying bar framework and its self-stress:

> **Theorem (Roth and Whiteley 1981).** Suppose that $T(G,P)$ is a tensegrity framework in $\mathbb{R}^d$. Then, the tensegrity $T$ is infinitesimally rigid in $\mathbb{R}^d$ if and only if the underlying bar framework $B(G,P)$ is infinitesimally rigid and $T(G,P)$ admits a proper self-stress that respects the typology of the elements.

The second result is that fusion (as well as adhesion) can be decomposed in a series of Type 1 and 2 Henneberg constructions (Figure B.1) which preserve the rigidity of a graph. In the case of a structure composed of two cells connected on three nodes, fusion and the removal of an edge can be seen as the addition of two nodes on a rigid bar framework and one edge removal. The resulting structure can be obtained by first adding a node through Type 1 Henneberg construction and then adding a second one through Type 2 Henneberg construction (which allows the removal of an edge). In the case of a structure composed of two cells connected on four nodes, fusion and the removal of an edge can be seen as the addition of a node on a rigid bar framework through a Type 2 Henneberg construction. In both cases, the resulting bar frameworks are thus rigid.

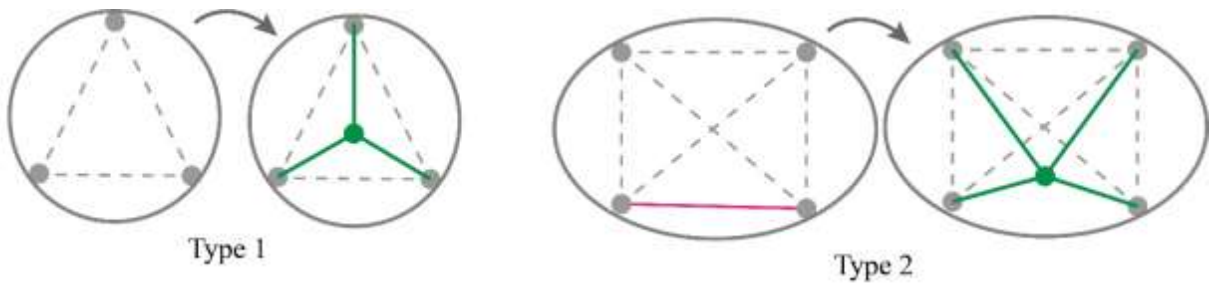

*Figure B.1: Type 1 and 2 Henneberg constructions in 3D.*

Since self-stress components dictate the element typology in cellular morphogenesis, the existence of a proper self-stress in the resulting bar frameworks is guaranteed which concludes the proof.

- Roth, B. and Whiteley, W., 1981. Tensegrity frameworks. Transactions of the American Mathematical Society, 265(2), 419-446.
- Tay T.S. and Whiteley W., 1985. Generating isostatic frameworks. Structural Topology 1985 Núm 11.

**Appendix C:**

A high cell-resolution bunny was generated using a tetrahedral mesh with a fifth node being added at the center of each tetrahedron to obtain Type II cell patterns (Figure C.1).

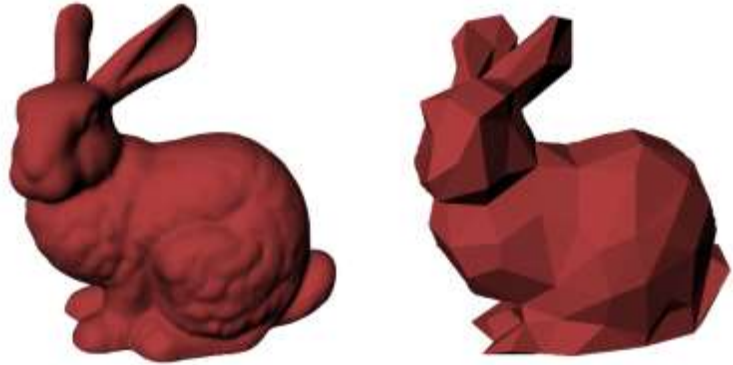

*Figure C.1: Illustration of the Stanford bunny structure (left) and the low polygonal surface mesh (right).*

The process resulted in 330 Type II cells with 528 nodes and 2126 members (806 elements in tension and 1320 elements in compression). The resulting structure (Figure C.2) has 548 self-stress states with 218 corresponding to virtual cells.

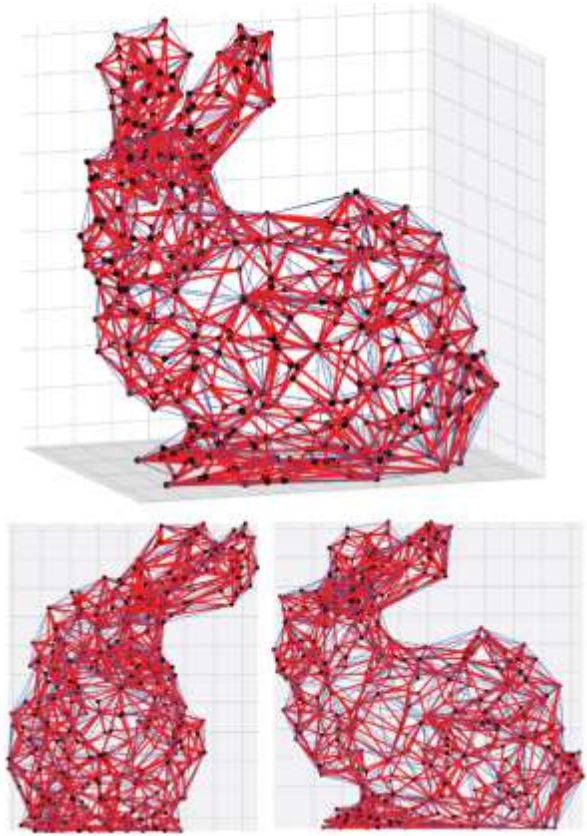

*Figure C.2: Illustration of the Stanford bunny tensegrity structure.*